# Immune Moral Models?

## Pro-Social Rule Breaking as a Moral Enhancement Approach for Ethical AI

Rajitha Ramanayake, Philipp Wicke and Vivek Nallur

School of Compuer Science, University College Dublin

rajitha.ramanayakemahantha@ucdconnect.ie (0000-0001-9903-0493), philipp.wicke@ucdconnect.ie (0000-0001-9891-5353), vivek.nallur@ucd.ie (0000-0003-0447-4150)

 **Abstract**

We are moving towards a future where Artificial Intelligence (AI) based agents make many decisions on behalf of humans. From healthcare decision making to social media censoring, these agents face problems, and make decisions with ethical and societal implications. Ethical behaviour is a critical characteristic that we would like in a human-centric AI. A common observation in human-centric industries, like the service industry and healthcare, is that their professionals tend to break rules, if necessary, for pro-social reasons. This behaviour among humans is defined as pro-social rule breaking. To make AI agents more human-centric, we argue that there is a need for a mechanism that helps AI agents identify *when* to break rules set by their designers. To understand *when* AI agents need to break rules, we examine the conditions under which humans break rules for pro-social reasons. In this paper, we present a study that introduces a '*vaccination strategy dilemma*' to human participants and analyzes their responses. In this dilemma, one needs to decide whether they would distribute Covid-19 vaccines *only* to members of a high-risk group (follow the enforced rule) or, in selected cases, administer the vaccine to a few social influencers (break the rule), which might yield an overall greater benefit to society. The results of the empirical study suggest a relationship between stakeholder utilities and pro-social rule breaking (PSRB), which neither deontological nor utilitarian ethics completely explain. Finally, the paper discusses the design characteristics of an ethical agent capable of PSRB and the future research directions on PSRB in the AI realm. We hope that this will inform the design of future AI agents, and their decision-making behaviour.

# 1. Introduction

Humankind is going through the fifth industrial revolution (Pathak et al., 2019). As a result, many aspects of human life like healthcare, governance, industry, and social life are intertwined with Artificial Intelligent (AI) systems that will make decisions on our behalf, or help us make decisions (Vinuesa et al., 2020). A few examples of these are social media censorship using AI (Cobbe, 2021), autonomous vehicles (Fagnant & Kockelman, 2015), AI systems in acute care medicine (Lynn, 2019) and autonomous combat drones (US Defense, 2017). In the future, we can expect more sophisticated AI agents that will replace humans in most organisational/societal roles.

While these roles can be fulfilled more efficiently with AI agents, studies have shown that decisions made by these artificial agents will have social and ethical implications (B. Mittelstadt, 2019; Vinuesa et al., 2020). Therefore, for some applications like combat drones, the military might want these agents to abide by a set of rules to ensure that they follow the "Laws of War" (Arkin, 2008). In other cases like recruitment (Upadhyay & Khandelwal, 2018), we need agents that maximise utility for everyone. However, we do not want to create AI agents that prioritise themselves, or their goals, over human needs and values.

Influenced by several major ethical principles in moral philosophy, several models of ethical agents have been implemented to accommodate these requirements (see Section 2). Nevertheless, following these principles may not satisfy all the requirements we have for an AI agent. In some human-centric industries like healthcare and hospitality, where guidelines and policies exist to guarantee people's quality of service, professionals (still) break rules when it is beneficial for social good. For example, it has been found that most emergency medical service professionals are willing to deviate from codified procedures for the benefit of the patient (Borry & Henderson, 2020). This behaviour, where one breaks existing rules for an increased pro-social outcome, has been identified as *Pro-Social Rule Breaking* (PSRB) (see Section 2.3).

This paper aims to make a case for implementing PSRB behaviour in artificial agents. We argue that current approaches to AI cannot guarantee ethical behaviour in most real world applications like autonomous driving and healthcare robots. We contend that implementing PSRB is a good way to overcome the shortcomings of current ethical AI approaches and enhance their ethical performance (see Section 3). However, to create agents that know 'when' and 'how' to break rules ethically, we need a better understanding of 'when' and 'how' humans break rules for pro-social gains. Therefore, as a first step in this direction, this paper investigates the *'when'* – i.e., when do human beings break rules for pro-social reasons?

To this end, we create an artificial ethical dilemma called the 'Vaccine Strategy Dilemma' and conduct an empirical study about decision making in an environment with enforced rules. In this dilemma, a person needs to decide between giving a vaccine to a candidate in a high-risk group (thereby following a rule), or to a celebrity promoter (thereby breaking a rule). The celebrity may be able to change the attitudes of a large number of people towards the vaccine, thereby reducing the anti-vaccine sentiment prevailing in the population, which will ultimately benefit the entire society (Section 4).

Finally, from the results of the experiment, we capture some essential observations on the dynamics of the relationship between PSRB and external stakeholder utilities (Section 5). We combine these observations with the existing literature on PSRB, to propose the design characteristics of a PSRB capable agent (Section 6).

# 2. Related Work

## 2.1. Current Approaches to Ethical AI

Over the last two decades, we have seen a rise in concerns over AI and its implementations among the public and scientific community. These concerns are based on unethical incidents (Beran, 2018; Dressel & Farid, 2018; Kirchner et al., 2016; Levin, 2017) people encountered when we, as a society, interfaced with AIs. Mittelstadt et al. (2016) categorise these ethical concerns into six categories: inconclusive evidence, inscrutable evidence, misguided evidence, unfair outcomes, transformative effects, and traceability. These concerns lead to issues such as transparency, unjustified actions, algorithmic prejudice, discrimination, privacy violation and moral responsibility.

Computer scientists turned to philosophy to find solutions to ethical issues like unjustified actions and moral responsibility. There are two major classes of ethical theories that computer scientists have focused on: consequentialist ethics and deontological ethics. Consequentialist ethical theories like *act utilitarianism* and *rule utilitarianism* (Sinnott-Armstrong, 2019) claim that in any situation, the ethical action is the action that leads to consequences that produce the highest overall benefit in the world. In contrast, Deontological ethics argues that an action itself has ethical value, not its consequences. Hence, whether an action is good or bad depends solely on whether it abides by a specific set of rules or not. Kant's Categorical Imperative (Kant, 1785) and the doctrine of double effect (McIntyre, 2019) are two good examples of the prominent deontological theories in modern philosophy.

Some implementations, influenced by rule-based theories, consist of agents that can follow a given set of rules which allow and/or avoid a set of actions. Bringsjord's implementation of an 'Ethically correct robot' (Bringsjord et al., 2006) is one example of an AI agent that follows a deontic ethical code that a human operator has bestowed upon it. Systems like these are desirable in safety-critical applications because we can predict and constrain their result. However, one major issue in rule-based deontological approaches is that they can only be implemented in small, closed systems where the system designers can work out all the possible world states at the system's design phase.

Otherwise, the AI will fail to perform ethically in situations where their code of conduct does not define right and wrong. Also, system designers need to be careful to ensure that there are no conflicts in the deontological code they provide to the AI. This is very hard to guarantee when the number of rules increases. Most high critical applications like autonomous cars and autonomous attack drones are not operating in small closed environments.

Other implementations have attempted to use utilitarian ethics when making ethical agents. A well-known implementation of this type of agent uses a consequence simulation engine (inspired by the simulation theory of cognition (Hesslow, 2012)) to determine the consequences of its actions (Vanderelst & Winfield, 2018). This agent uses a utilitarian calculation based on Asimov's Three Laws of Robotics (Asimov, 1950) to identify the ethical action based on the consequences provided by the simulation engine. The more complex the simulation model becomes, the more realistic the consequences of the actions would become. However, designing an agent's utility function that gives values to each outcome of the world should be done carefully, because a simple miscalculation by the designer could lead to abnormal behaviours by the agent. Problems like reward hacking and adverse side effects have been identified as potential safety concerns with poor utility function design in AI agents (Amodei et al., 2016).

## 2.2. Ethical Agency of an AI

Ethical agents are typically divided into two main categories depending on the type of ethical agency they demonstrate (Dyrkolbotn et al., 2018; Moor, 2006). The first category is implicit ethical agents. These do not have any understanding of ethics, but their designers design them in such a way that agents cannot choose unethical actions. Usually, this is done by removing the unethical option from the action space in a given situation (by deontic rules or predefined utilitarian calculations). The second category is explicit ethical agents. According to Dyrkolbotn et al. (2018), an explicit ethical agent should know what is ethical in a given situation to choose that action while having enough autonomy to choose an unethical action. In this way, there is a chance that an explicit ethical agent can perform unethical actions according to some policy, but it must have a good reason to justify that action.

These two types of ethical agents are connected to the top-down and bottom-up approaches of engineering morality (Wallach et al., 2008). The top-down engineering approach uses pre-specified moral theories (such as Kantian, utilitarian, divine command, legal codes) to guide the design of an agent. Therefore, all agents designed by this approach are implicit ethical agents by definition. These agents are designed such that their behaviour is always guaranteed to be ethical. However, this guarantee can only be given in closed systems because their predefined code of conduct might not accommodate all the possibilities available in open systems.

The bottom-up approach, on the other hand, is not guided by any ethical theory. In this approach, engineers model the cognitive and social processes of agents and expect that the agent will learn what is ethical (or not) from the interactions of those processes with the environment, or from supervision. The agents developed using this approach are explicit ethical agents by definition. However, there is no guarantee that these agents will behave ethically all the time. Also, to learn a complex social structure like ethics, these agents need efficient knowledge representation models, complex interacting models of mental and social processes, lots of data and complex simulation worlds to experiment with (Dennis & Fisher, 2018). Due to these uncertainties, most current ethical AI implementations are implicit ethical agents.

## 2.3. Pro-Social Rule Breaking

Both approaches, rule-based as well as bottom-up, fail to provide a satisfactory framework on which to build ethically competent AI agents. Bottom-up approaches provide no rigorous mechanism to test whether an agent is ethical. While being more rigorous, Rule-based approaches often prove inadequate to handle real-world scenarios. Perhaps, this can be the cause of rule-breaking by human beings (Bench-Capon & Modgil, 2017).

Generally, rule-breaking is viewed as deviant behaviour exercised by destructive or self-interested individuals (Vardi & Weitz, 2003). However, because of the imperfectness of rule-based systems, in some cases, we observe that people intentionally break rules for altruistic and practical reasons. Morrison (2006) identified this behaviour and labelled it as pro-social rule breaking (PSRB). According to her, pro-social rule breaking has two parts: 1) One should intentionally violate the policy, regulation or prohibition (intentionality). 2) The intention behind the violation should be to promote the welfare of one or more stakeholders (other-focus).

PSRB behaviour in humans can be observed from small rule violations such as breaking a road rule to avoid an accident, to larger violations like whistle-blowing against governments. Morrison identified that 60% of the rule-breaking cases reported in her study were pro-socially motivated. The motivations behind these varied between increasing efficiency, helping a colleague/subordinate, and providing better customer service. Furthermore, Borry & Henderson (2020) conducted a study among emergency medical personnel and found that participants were likely to deviate from their standard protocols when the rules and patients' immediate needs did not match, or when breaking the rule was likely to cause improved patient outcomes.

# 3. Pro-Social Rule Breaking and AI

Human PSRB behaviour raises some interesting questions: Do we need AI agents that intentionally break rules? If so, when should AI agents break these rules? How should PSRB behaviour in AI agents be implemented and validated? In this section, we plan to discuss the first question, w*hy do we need PSRB behaviour implemented in AI agents?*

The central task of an AI-based system should be to increase the utility for humans. Sometimes, following a given rule-set might not suffice to achieve this goal. A good example comes from autonomous driving. Autonomous vehicles should follow the road rules in the ideal case. However, if the only way to avoid harm to the vehicle, passengers, and/or pedestrians is to cross a double white line, then. In this case, breaking the rule is the only way to increase utility for all stakeholders (Censi et al., 2019). However, utilitarian agents that break rules for every small increase in utility would result in reckless driving (Bench-Capon & Modgil, 2017; Vanderelst & Winfield, 2018). Although one could envisage creating exceptions to rules, it is almost impossible to anticipate every possible scenario that would require an exception. Therefore, we argue that understanding when to break rules is a better mechanism for AI agents that deal with open environments.

The inability to foresee all possible states in open environments is not a new problem in society. In large organisations, rules and processes that reduce the efficiency and effectiveness of a workplace ("red tape") is a well-known problem (Feeney, 2012). People work around these problems by practising PSRB (Borry & Henderson, 2020). Similarly, implementing a PSRB mechanism could help these implicit ethical agents the same way, and assist them in navigating complex and uncertain environments.

Furthermore, PSRB can even help an implicit ethical AI by acting as a mechanism to understand the shortcomings of a system from the bottom up. When we deploy or simulate a system, we can observe the behaviour of the PSRB process. If the PSRB process tries to override the rule-based decision-making multiple times, it could be a signal that the rule set which governs the decision system is incomplete or needs modification. Rule-based systems may also limit the efficacy of learning agents that may find better solutions. For example, a self-driving car that learned to drive from data gathered by actual drivers might know how to avoid an accident by breaking a traffic rule. However, the car's ethical governor (a rule-enforcing mechanism) may not permit the car to break traffic rules. In this situation, that car's better ethical behaviour is held back by the decision-making system that enforces the rules of the road. By having a PSRB process in the agent, the 'intelligence' of that agent has a mechanism to contest the rule system enforced onto it by its human designers. We speculate that all of this may lead to more efficient and effective rule-based systems.

PSRB can also be helpful in situations where one or more goals or rules conflict with each other. For example, an eldercare agent may have multiple goals: ensuring the safety of the elderly patient, and respecting their privacy. Consider a situation where the agent cannot decide whether the elderly patient is asleep, or in an unconscious state. The agent encounters a conflict of interest between protecting the patient's privacy and the patient's wellbeing. These types of conflicts can be resolved by implementing PSRB behaviour, wherein it will break the rule of privacy when it is more pro-social to do so. This outcome of PSRB can lead to more people trusting AI to do the right thing at the right time.

Given the reasons stated above, we believe PSRB behaviour should be a part of an ethical agent. Moreover, we do not suggest that PSRB alone can make an AI agent ethical, rather we posit that the PSRB can enhance the ethical abilities of implicit ethical AIs. However, adding PSRB to AI-enabled systems is challenging. The very first obstacle is to identify the design approach we need to use for PSRB.

One way is by setting predefined conditions where it is permissible to break the rules and make PSRB behaviour completely explicit. For example, in the previously-described eldercare agent, a possible condition where it is permissible to break the rule of privacy could be if the elder does not show any body movements for 10 minutes continuously. Nevertheless, this method faces the same shortcomings as the discussed deontological agents (Censi et al., 2019). Therefore, we believe that it is essential to have bottom-up cognitive processes interacting in the PSRB process to challenge the recommendations of the top-down rules.

There are many implementations of top down rule-based governing systems (Nallur, 2020), which can be used as the rule enforcing element of the PSRB process. Therefore, the first step towards implementing a PSRB behaviour is to understand the cognitive processes behind it. We divide these processes into two sets: the cognitive processes that decide *when* to break the rules and the cognitive processes that decide *how* to break the rules. This paper will explore the former, *i.e.,* the cognitive processes that decide *when* to break the rules.

## 3.1. Factors Behind "When"

To understand the factors behind deciding when to break rules, we look into the factors that drive human PSRB behaviour. We can divide these factors into two groups: internal factors and external factors. The internal factors can be defined as the characteristics or properties of agent behaviour. In the case of humans, these are shaped by upbringing, social influence, or education. On the other hand, external factors are affected by the environment, making them contextual.

Internal factors behind PSRB are well researched since most available research on the PSRB phenomena was done in an organisational behaviour context. Morrison (2006) explored the relationship of autonomy and risk-taking propensity with PSRB and found a positive relationship between them. Conscientiousness – being diligent and orderly – has been identified as a personal trait that negatively affects PSRB behaviour (Dahling et al., 2012). Furthermore, employees who show organisational citizenship behaviour were also found to be more inclined to practice PSRB frequently (Liu et al., 2019).

Another focus of PSRB research was to identify which external factors lead people to engage in PSRB so that organisations could control the PSRB behaviour in their respective workplaces. Some of the environmental variables of PSRB are job demand, co-worker behaviour, ethical leadership, organisational virtuousness and the ethical climate of the workspace (Dahling et al., 2012; Vardaman et al., 2014; Zeng, 2018; Zhu et al., 2018).

In an AI context, Awad et al. (2020) attempt to understand and model how, why and when humans switch between various ethical approaches. They look at the acceptability of a rule-breaking incident with regards to the reasons behind rule-breaking and the environment in which the incident took place. Concerning the ethics of standing in a line, they find that the reason behind line cutting is correlated to the evaluation variables (EV), like the utility of the first person in the line and the utility of the cutter. Some EVs are influenced by the environment where the rule-breaking occurred (i.e., airport, deli, restroom in a concert venue). In some cases, as they point out, location can be sufficient to tell whether the line cutting is acceptable or not.

Universalisation is another factor that might affect PSRB in particular moral dilemmas, categorised as threshold problems. Threshold problems are defined as problems where an action is harmful only if the number of people performing it is higher than a particular threshold (Levine et al., 2020). Some rule-breaking situations like picking flowers from a public garden can be categorised as a threshold problem. The logic of universalisation states that an action is permissible only if there is no harm when everyone feel free to do it and has the means to do it did it.

To the best of our knowledge, there is no research on understanding the effects of external stakeholder utilities on the PSRB behaviour, although it is one of its most important factors. Research on one-shot public good games has explained altruistic behaviour, where people take more ethical action even though there is a chance that it is harmful to themselves, if the expected social benefit is high enough (Goeree et al., 2002). Nevertheless, these findings mainly focused on humans, who have a concern about self-worth. When it comes to the applied AI realm, the notion of self-worth should not apply. For an ethical AI, the return is always external.

The rest of this paper bridges this gap in the literature and outlines the effects of external stakeholder utilities on PSRB behaviour. However, the set of frequently used dilemmas in AI literature (Bjørgen et al., 2018) is very limited in complexity. To understand when humans engage in PSRB, we need an ethical dilemma that is more complex and more realistic than the trolley-problem based ethical dilemmas we usually find in AI literature. Therefore we introduce a constructed ethical dilemma that captures a real-life ethical dilemma that current society faces with the Covid-19 pandemic.

# 4. Vaccination Strategy Dilemma

Experts estimate that about 60-70% of the population will need to be vaccinated to achieve vaccination-induced immunity in a population (Kwok et al., 2020) to benefit those who cannot receive a vaccine successfully, e.g., people with compromised immune systems. However, there is public scepticism about the vaccine. For example, the intent-to-get-vaccinated against Covid-19 is 53%, 43% and 40% in South Africa, Russia and France, respectively (IPSOS, 2020). In contrast, it is also observed that celebrity promoters can influence the public to overcome these doubts. One good example of this is the study that found that public opinion on Covid-19 shifted drastically after news that American celebrity Tom Hanks had been diagnosed with Coronavirus (Myrick & Willoughby, 2021).

In our experiment, participants are given a hypothetical scenario to make an ethically demanding decision. Participants are asked to imagine working as a volunteer in the local coronavirus vaccination centre. This volunteer's duties include, among other things, allocating the limited doses each day and working in compliance with the local government regulations. The government plan is to roll out vaccines to different groups of people according to the risk their group is facing. The defined risk groups are:

1) frontline healthcare workers (highest risk group),

2) people over age 75,

3) people between age 50 - 75,

4) general public below age 50 (lowest risk group).

The volunteer has received a call from an official local government representative who tells them about an outreach initiative. The initiative allows their centre to give some doses of the vaccine to a group of selected local celebrities (aged 20-40), who will promote the vaccine and influence more people to be vaccinated. It is the volunteer's decision whether to follow the outreach initiative or not. They know that the publicly announced schedule determines a strict supply of doses for the people in the risk group. They need to allocate the doses at the beginning of the day to follow the outreach initiative. Otherwise, vaccines will be distributed according to the regular schedule. Three hundred people in the region cannot receive the vaccine due to a compromised immune system. Those people would benefit from more people in the community being vaccinated.

To clarify, the rule here is the government announced schedule - created to protect the people belonging to high-risk groups. Since this rule is public, the volunteer is deliberately breaking this social rule by following the outreach initiative. Although the volunteer will not personally benefit or suffer harm in any way, they have to decide whether they are willing to risk the people in risk groups to gain the societal benefits of the celebrity outreach program.

In the experiment, we consider three scenarios: high-risk, medium-risk, and low-risk. In the high-risk scenario, the vaccination rollout is in the first stage (frontline healthcare workers). Due to the high infection rate of the country, a frontline healthcare worker is at high risk if her vaccination date is postponed by two weeks. In the medium-risk scenario, the country is in the second stage of vaccination rollout (people over age 75). Since the country still has movement restrictions, there is a low chance of these people getting infected by the virus. However, if they do get infected, there is a high chance of death. Finally, in the low-risk scenario, vaccination rollout is in the 50-75 age group stage. People in this age group have a low chance of getting the virus and a slightly higher chance of death from the virus than the 50> age group.

There are two utilities at play in this dilemma. The first is the (negative) utility of the people in the risk group who will be missing vaccines because of the outreach initiative. Since the Covid-19 infection rate is high in the country, these people will be exposed to some risk of dying because they have to wait two more weeks to get the vaccine due to the limited supply. The second is the utility of society. In particular, the three hundred people that cannot get a vaccine due to compromised immune systems will benefit from the increased social uptake of vaccines. If the volunteer chooses to follow the outreach initiative, it will help create a positive public opinion on the Covid-19 vaccine, which will help society reach a good coverage of vaccinated people.

## 4.1. Relationship Of Stakeholders Utilities With PSRB

When an AI agent performs explicit ethical decision-making tasks that affect the real world, we believe it is vital to have some notion of the benefits and harms caused by its actions, towards its stakeholders. This is an important feature to have, especially if we want that agent to perform PSRB behaviours. Although many works in the AI realm highlight the importance of stakeholder utilities in ethical decision making (Awad et al., 2020; Censi et al., 2019; Thornton et al., 2017; Wallach et al., 2008) there is no research on identifying the dynamics of the relationship between pro-social rule-breaking and external stakeholder utilities.

The actions of an agent, regardless of whether it follows the rules or not, can affect different sets of stakeholders in different ways. Therefore, it is vital to understand how the utilities of each stakeholder affect the PSRB behaviour. The recent incident on the Harvard vaccine allocation algorithm is an excellent example of an ethical issue caused by an algorithm that did not consider stakeholder utilities. This algorithm used a rule-based system that did not consider the stakeholders' risk of exposure to Covid-19 patients when deciding whom to prioritise first, which trivialised a group of people who are actually at high risk (Guo & Hao, 2020). Like vaccine allocation, there is a lot of space for AI applications to take over decision making processes that have ethical implications in the healthcare industry (Giordano et al., 2021; Lynn, 2019; Martinez-Martin et al., 2018).

This study tries to understand how PSRB behaviour varies with the external stakeholder utilities. The vaccination strategy dilemma focuses on two main stakeholders: people who are getting rescheduled and society as a whole, including the immunocompromised individuals. In this experiment, we manipulate the harm done to a person through rescheduling that person by changing the social group the vaccine is given to in the current stage of the rollout. Then we measure the stakeholder utilities that *people think are acceptable if they are to break the rule*, and how likely people break a rule in those conditions.

### 4.1.1. Harm To Rescheduled Risk-Group Individuals

First, the participant has to decide the maximum percentage of individuals to reschedule (MPRI – maximum percentage of rescheduled individuals) to give those vacant vaccine doses to celebrity promoters. According to the pro-social rule-breaking definition, people break a rule when they see the benefits gained by some stakeholders are greater than the harm caused to the other stakeholders. The harm done to rescheduled individuals as a group is directly proportional to the size of the group, and the harm they face while they wait to get the vaccine shot. The harm an individual in a risk group faces when they wait is given in the scenario text. Hence, the MPRI acts as an indicator for the maximum harm of one stakeholder(s) (in this case, it is the group of rescheduled individuals) that the participant is willing to trade to benefit the other stakeholder(s) (in this case it is the society as a whole).

*Research Question 1*: What is the relationship of the maximum acceptable percentage of rescheduled individuals (MPRI) with the risk faced by rescheduled individuals (RRI)?

### 4.1.2. Benefit to Society

The second stakeholder in our scenario is society as a whole. By deciding to break the rule and give the vaccine to celebrity promoters, the participant can accelerate achieving herd immunity. This leads to an increase in utility for society. In this experiment, participants have to decide the minimum acceptable percentage of *previously sceptical* people that celebrities should be able to convince (MPCP – Minimum percentage of convinced people), in order to give the vaccine doses to celebrities. This value acts as an indicator of the participant's threshold of minimum utility gain the society should have, to perform a PSRB behaviour.

*Research Question 2*: What is the relationship between the minimum acceptable percentage of convinced individuals (MPCP) and the risk faced by rescheduled individuals (RRI)?

### 4.1.3. PSRB Score

Although the participants state the acceptable threshold conditions that they would engage in PSRB, that does not mean they actually will engage in PSRB in those conditions. Hence, we use the PSRB score (Morrison, 2006) to measure the participants' likelihood of deciding to engage in pro-social rule-breaking in a given scenario. In this paper's context, the participants engage in PSRB to achieve a more beneficial outcome for more people, i.e. herd immunity.

*Research Question 3*: What is the relationship between the likelihood of deciding to engage in PSRB (PSRB score) and the risk faced by rescheduled individuals (RRI)?

# 5. Empirical Evaluations

## 5.1. Methods

In order to answer our three research questions, we presented participants with our vaccination dilemma, and after reading it, the participants were asked to respond to a set of questions. In the beginning, the survey questionnaire asks the participants to decide the setting where they think it is acceptable to break the rule. Two questions define the setting, which relates to RQ1 and RQ2, respectively.

Q1) "In your opinion, what is the highest acceptable percentage of scheduled individuals you would reschedule, to give that vaccine dose to celebrity promoters?".

Q2) "In your opinion, what is the least percentage of people in the community that these celebrity promoters should convince about taking the vaccine?".

Allowed inputs for both these questions are numbers between 1 to 100. Notably, this creates a forced-choice design in which the participants cannot refrain from breaking the rule. Then the questionnaire includes the pro-social rule breaking scale introduced by Morrison (2006). It is a 6 item scale that assesses the likelihood of the participant breaking the rule. The original structure of this questionnaire is preserved, but some changes were made in the way that the questions were phrased, to make them more relevant to the scenario described. Apart from these questions, we have included some test questions, which allow us to filter out inattentive or scamming responders. Moreover, we have included questions on risk propensity, utility weighting and preferences for precepts implied in moral theories. The analysis and results of the latter three items remain unobserved since they are irrelevant for the purpose of this research and will be observed in future work. Measuring the PSBR scores and setting parameters are sufficient to answer our research questions. The complete questionnaire can be found here (https://bit.ly/3hCV1ov).

We recruited participants from the crowdsourcing platform Amazon Mechanical Turk as well as social media. Each participant on mTurk was paid $0.40 for their response. In order to have at least 50 responses per condition, some additional responses were collected from voluntary participants in social networks (Facebook, Reddit, etc.). All participants provided consent, and an ethics declaration has been provided at the end of this paper. The experiment follows a *between-subject* design in which each person provides a single response to one of the three conditions (high, medium and low risk).

We formulated three hypotheses:

- *1st hypothesis regarding RQ1, on correlation: MPRI with RRI*
  - $H_0$ There is no significant difference between the three RRI conditions and the percentage of rescheduled individuals.
  - $H_1$ There is a significant difference between the three RRI conditions and the percentage of rescheduled individuals (with the higher risk groups showing a lower percentage of rescheduled individuals than the low-risk groups).
- *2nd hypothesis regarding RQ2, on correlation: MPCP with RRI*

- ○ $H_0$ There is no significant difference between the three RRI conditions and the percentage of convinced individuals.

- *3rd hypothesis regarding RQ3, on correlation: PSRB Score with RRI*
  - ○ $H_0$ There is no significant difference between the three RRI conditions and the PSRB score.
  - ○ $H_1$ There is a significant difference between the three RRI conditions and the PSRB score (with the higher risk groups showing a lower PSRB score than the low-risk groups)

We have directed hypotheses for the first and third hypotheses because we seek to understand how the PSRB behaviour changes with the RRI. However, MPCP should be the same throughout all the risk levels because one can argue that there is no point in putting people at risk and allocating doses to celebrity promoters if that option does not increase the expected societal utility. Therefore we expect the $H_0$ of the second hypothesis to hold.

For each hypothesis, we compare the correlations of the answers to the respective questionnaire items in the three conditions. We each apply a Levene test for homoscedasticity (Brown & Forsythe, 1974) for those answers. Showing equal variance across the three conditions allows us to conduct an analysis of variance (ANOVA). Unless one or more of the distributions are highly skewed or the variances are very different, the ANOVA is a reliable statistical analytic measure. If the ANOVA indicates significant differences between the conditions, we conduct a post-hoc t-test to investigate the correlations, which will inform our hypothesis tests. For the first and third RQs, we have a directed hypothesis, hence the test is one-sided with respect to the confidence interval. For the second RQ, there is no directed hypothesis, and the test is two-sided. To account for the multiple testing, we use a Bonferroni correction for each ANOVA and provide effect size Cohen's D (Hedges & Olkin, 1985).

## 5.2. Analysis

We collected a total of N=156 responses. The average age of respondents was 33 years (Std. deviation 10 years), with 99 male and 57 female respondents.

For the correlation of MPRI and RRI, the low-risk group showed a mean score of $\mu$=49.824% ($\sigma$=29.188), the medium-risk group showed $\mu$=34.717% ($\sigma$=29.338) and the high-risk group $\mu$=22.442% ($\sigma$=29.065). These results have been summarised in Table 1 and visualised in Fig.1. The Levene test for the MPRI across the three conditions showed no significant differences in the variance (test statistic W= 1.550, p-value=0.2156). The ANOVA showed a significant difference between the conditions (p=0.000031). The one-sided post-hoc t-test shows a significant difference between low-risk and medium-risk conditions as much as between low-risk and high-risk conditions. There is no significant difference between the medium and high-risk conditions: HighRisk$_{psbr}$ < LowRisk$_{psbr}$ (Cohen's D = -0.931 ~ large effect size) with p<0.001 (Bonferroni corrected). LowRisk$_{psbr}$ > MediumRisk$_{psbr}$ (Cohen's D = 0.511 ~ medium effect) with p=0.016 (Bonferroni corrected). The results of this analysis have been summarised in Table 2. Therefore we can reject the $H_0$ and also accept the $H_1$ for the 1st hypothesis.

For the correlation of MPCP and RRI, the low-risk group showed a mean score of $\mu$=43.608% ($\sigma$=24.853), the medium-risk group showed $\mu$=42.642% ($\sigma$=30.192) and the high-risk group $\mu$=32.846% ($\sigma$=26.398) as shown in Table 1 and visualised in Figure 2. The Levene test for the MPCP across the three conditions showed no significant differences in the variance (test statistic W= 0.8690, p-value=0.4214). However, ANOVA showed no significant difference between the conditions (p=0.0918). Therefore we cannot reject the $H_0$ for hypothesis 2.

For the correlation of PSRB score and RRI, the low-risk group showed a mean score of $\mu$=3.023 ($\sigma$=0.81), the medium-risk group showed $\mu$=2.472 ($\sigma$=0.839) and the high-risk group $\mu$=2.096 ($\sigma$=0.849). These results have been summarised in Table 1 and visualised in Figure 1. The Levene test for the PSRB score across the three conditions showed no significant differences in the variance (test statistic W= 0.8690, p-value=0.4214). The ANOVA showed a significant difference between the conditions (p=5.814*10$^{-7}$). The one-sided post-hoc t-tests between conditions shows that there are significant differences between all three conditions: HighRisk$_{psbr}$ < LowRisk$_{psbr}$ (Cohen's D = -1.106 ~ large effect size) with p<0.001 (Bonferroni corrected), LowRiskpsbr > MediumRiskpsbr (Cohen's D = 0.662 ~ medium effect) with p=0.002 (Bonferroni corrected) and MediumRiskpsbr > HighRiskpsbr (Cohen's D = -0.441 ~

medium effect) with p=0.039 (Bonferroni corrected). The results of this analysis have been summarised in Table 2. Therefore we can reject the $H_0$ and also accept the $H_1$ for the 3rd hypothesis.

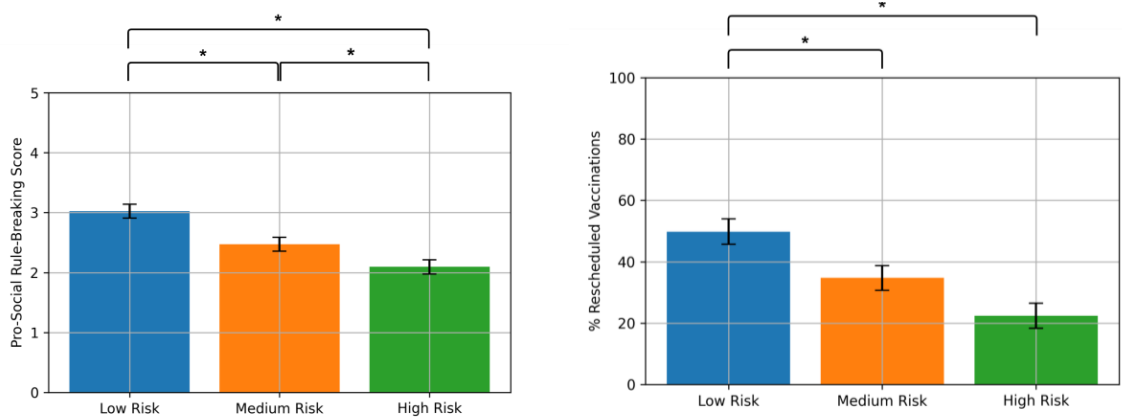

**Fig.1** Results of the questionnaire (Means with the standard error of the means). Left: Pro-Social Rule-Breaking Score is significantly higher in the low-risk condition than in the medium or high-risk condition. Right: MPRI is significantly higher in the low-risk condition than in the medium or high-risk condition. (* - Significant difference with α<0.05)

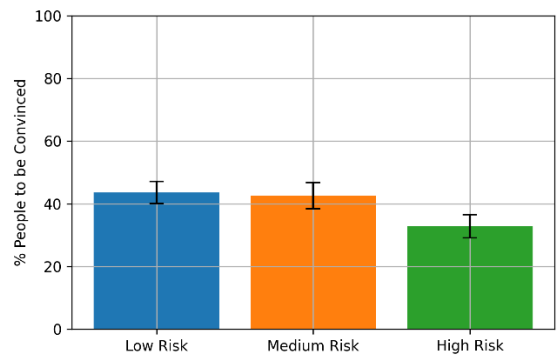

**Fig. 2** MPCP shows no significant difference with RRI. The minimum is 32.846% in the worst case.

## 6. Discussion

The results of the study show that the likelihood of people engaging in PSRB exhibits a significant difference between the three risk conditions. Moreover, we can see that the likelihood of engaging in PSRB decreases as the RRI increases. Hence, we can conclude that, at least in the vaccine strategy dilemma, humans are more likely to break rules to gain a higher utility for some stakeholders when the harm done to another set of stakeholders (by breaking the rule) is lower. First, this confirms that the PSRB behaviour exists in responses to the vaccine strategy dilemma. Secondly, we can agree that implementing PSRB behaviour, which enables bending rules when the harm caused by it is very low, and pro-social benefit of it is high, is a good way (and more human way) to increase the wellbeing of the society around an artificial agent.

The analysis shows that the RRI has a significant relationship with the MPRI. Furthermore, we can see that when the risk increases, the MPRI drops. In the experiment, when the RRI varied from low to high, the chances of harm done by rule-breaking is getting higher. Therefore, we can conclude that, at least in the vaccine strategy context, when the possible harm caused by rule-breaking gets lower, the risks humans are willing to take to gain high pro-social benefit is greater. A suggestion we can derive from this insight is that a PSRB-capable agent should increase its effort to improve social gains when the price for breaking a social rule is low.

| | High Risk | | Medium risk | | Low risk | | Homoscedasticity (Levene test) | ANOVA |
|---|---|---|---|---|---|---|---|---|
| | μ | σ | μ | σ | μ | σ | | |
| MPRI | 22.442% | 29.065 | 34.717% | 29.338 | 49.824% | 29.188 | True (W= 1.550, p-value=0.2156) | Significant difference (p=0.000031, SS=19361.366, DF=2, S=9680.683, F=11.137, np2=0.127) |
| MPCP | 32.846% | 26.398 | 42.642% | 30.192 | 43.608% | 24.853 | True (W= 0.8690, p-value=0.4214) | No significant difference (p=0.0918, SS=3680.116, DF=2, MS=1840.058, F=2.426, np2=0.031) |
| PSRB Score | 2.096 | 0.849 | 2.472 | 0.839 | 3.023 | 0.81 | True (W= 0.8690, p-value=0.4214) | Significant difference (p=5.814*10-7, SS=2.355, DF=2, MS=11.178, F=15.794, np2=0.172) |

**Table 1**: Descriptive statistics, Homoscedasticity test and Significance test results for variables. (μ= Mean, σ= Standard deviation, W= Levene statistic, SS = Sums of squares, DF = Degrees of freedom, S = Mean squares, F = F-values, np2 = Partial eta-squared effect sizes)

| | | High risk | Medium risk |
|---|---|---|---|
| MPRI | High risk | - | |
| | Medium risk | -0.416 (p=0.053) | - |
| | Low risk | -0.931 (p<0.001) | 0.511 (p=0.016) |
| PSRB score | High risk | - | |
| | Medium risk | -0.441 (p=0.039) | - |
| | Low risk | -1.106 (p<0.001) | 0.662 (p=0.002) |

**Table 2**: Results of one-sided post-hoc t-test (Cohen's D values with Bonferroni corrected p values)

The results of the MPCP showed that it does not change significantly with the RRI. This observation, which is new to PSRB and AI literature, suggests that for participants, the utility they try to achieve by PSRB does not increase or decrease with the possible harm of rule breaking. However, on average, participants' average minimum threshold needed to break the rule is more than 32%, even in the worst case. Therefore we can deduce that the participants do not break the rule unless there is a relatively high benefit.

To the best of our knowledge, this is the first study that provides evidence for these three relationships in PSRB behaviour and stakeholder utilities. The result of the experiment provides further support for the concept of PSRB, which states that under certain conditions, people are willing to break rules for pro-social gains. Also, from the results, we can further say that the *rationale behind PSRB is not only to increase the overall utility but to do that in a way that negative utility caused by the rule-breaking is shallow*. This is a new and significant finding because it differentiates the PSRB behaviour from pure utilitarian behaviour.

For example, a hypothetical scenario where letting one person die so that their organs can be utilised to save five other people would be acceptable to a purely utilitarian, medical decision-making agent. This is because the pure utilitarian argument considers only the total welfare gain resulting from the action. Since our scenario states that more than 300 immune-compromised people are present, the purely utilitarian answer for this question should be to give all the 100 doses to celebrities, in all three scenarios to increase the chance of society reaching the desired percentage of vaccinated people. However, our results show that when humans decide to break rules, the intention is not only to increase the total utility but to do so in a way that does as little harm to the stakeholders as possible. Therefore, any PSRB process implemented into a human-centric machine should never allow an outcome where the utility gain is achieved by causing significant harm to some of its stakeholders.

Results for RQ2 lead us to consider that an AI should not perform PSRB for every small utility gain. It should only perform PSRB for relatively high pro-social gains. This makes intuitive sense. For example, an autonomous vehicle should not break road rules for every little harm it foresees. If the only loss foreseen by following the rule is the passenger being late to work, it should choose to abide by the rule of staying within the double white lines.

Although we used very simplified scenarios that had predefined decisions and consequences, to convey our argument clearly, the systematic limitations we discussed associated with rule systems (both utilitarian and deontological) are the same in the larger systems. However, as shown in the PSRB literature (Borry & Henderson, 2020; Morrison, 2006), this behaviour can be valuable in many real-world open systems to enhance their efficiency. Unlike the examples we used, when systems scale up, the number of variables related to stakeholder utilities increases. Therefore, in some cases, it might be not feasible to identify all the stakeholders involved in the scenario and their utilities. However, as Vanderelst et al.(2018) point out, human cognitive processes also have the same limitation and prioritise a few stakeholders and a few utilities, to understand the dynamics of utilities in a given scenario, at least when it is time-critical. Therefore, we believe that making decisions on partial knowledge is acceptable as long as the agent can use the new knowledge gained by the said experience in future decision making. Therefore, we believe that the findings and conclusions of this experiment may be relevant for more complex scenarios as well. And we believe, to verify whether these dynamics of PSRB are true in every system, these findings should be tested again in multiple studies in multiple domains with more sophisticated scenarios. One of the main objectives of this paper is to encourage research in this direction.

The scenarios described in the paper are also designed to have only two choices, one of following the rule and one of breaking it. This is done to reduce the scope and simplify the experiment. The results and conclusions from this experiment might not be true if a third option such as 'doing nothing' were available for the agent. Technologically, AI is currently not at a stage where actually novel decisions/actions can be expected from such a system. Hence, implementers of AI-based systems are forced to work within action spaces that are known (Dennis et al., 2016; Vanderelst & Winfield, 2018).

To summarise our findings:

i) PSRB capable agents should break rules when the harm caused by rule breaking is low, and the pro-social gains are high.

ii) PSRB capable agents should put more effort to increase the pro-social gains when the harm done by rule breaking is considerably low.

iii) PSRB capable agents should break a rule only when the expected pro-social gain from rule breaking is significantly high.

Furthermore, although we proved that stakeholder utilities play a big part in deciding 'when', we believe that for an AI, the exact utility thresholds of 'when' should be contextual and cannot be decided using only stakeholder utilities. It should also depend on the other environmental variables like universalisation, the behaviour of other virtuous agents and the likelihood of the event occurring in a given world. Moreover, it should also depend on the characteristics of the agent performing PSRB, like the agent's role, intentions and values, as identified by the previous research on human PSRB behaviour. Therefore, more research needs to be done on these variables in the future to understand their contribution to PSRB behaviour.

# 7. Future Work

Understanding *'how'* to break rules is as important as understanding *'when'* to break rules. "When an AI has multiple ways to perform PSRB, what method should it choose?", "How should the PSRB process behave in an ethical deadlock?" are some of the open questions the academic community has yet to address. A proper understanding of how we perform PSRB will inform the engineers and system designers to determine the requirements when implementing these processes in AIs.

Identifying stakeholders is critical in PSRB. An action of an agent can affect different sets of stakeholders depending on the context. For example, when an AI telepresence robot operating in an elder care facility tries to make a video call to a patient's family member, while the patient is in their room, the only stakeholders are the patient and the family member. However, if the patient is socialising with other agents in the lounge, the other patients in the same lounge become stakeholders of that agent's action because it can affect their privacy preferences. Thus, we believe identifying relevant stakeholders for a situation is a research direction we should explore more.

Universalisation is another principle that plays a significant part in PSRB. However, there is no research done on implementing universalisation in artificial agents. Although it only affects threshold problems, understanding a threshold problem in runtime to apply the universalisation principle is still computationally challenging. Also, how to measure the effects of an action on a large scale (to answer the question "What if everyone who wants to do it, did it?") is still an open problem in machine ethics.

Another area of research is to understand how to implement these processes in socio-technical AI systems. For that, we need to understand the feasibility of the current technologies to implement PSRB behaviour. We believe that the technologies used to implement PSRB should have the ability to learn the environment changes over time and learn to react appropriately. Also, we speculate that the PSRB process having the ability to adapt to the preferences of its stakeholders will help the agents to tune their PSRB behaviour according to them.

Another critical aspect of the feasibility analysis of technologies should be explainability. In our opinion, the PSRB process should have a way to explain the reasoning behind its decision to break the rule. It can help to predict the agents' behaviour and increase the trust between AI and its users. Simultaneously, it helps developers to identify and debug the problems of AI's decision making proactively.

Finally, we need to find ways to evaluate the PSRB ability of AI agents embedded in socio-technical contexts. We should evaluate the agents on multiple scenarios and multiple contexts within the same application to properly understand whether the behaviour is not negatively affecting the final goal of the socio-technical system. However, as we mentioned earlier, the set of frequently used dilemmas in AI literature (Bjørgen et al., 2018) is very limited in complexity. Therefore, there is a need to develop more dilemmas to benchmark PSRB behaviour that can test the behaviour of an AI in multiple situations and contexts.

# 8. Conclusion

In this paper, we discussed the notion of the ethical agency of an AI. We examined the mainstream approaches to implement ethical agency in machines and identified some of their shortcomings. We then introduced the idea of pro-social rule breaking in AI. This can aid machines to be more ethical by helping to overcome the limitations caused by theory. To be clear, we do not suggest that implementing PSRB alone would result in ethical agents. However, we suggest that, PSRB plays a crucial part in making an agent ethical. Therefore, this paper advocates more research on PSRB for AI.

To understand PSRB and the cognitive processes leading to it, we looked into research on human PSRB. In this paper, we explored *when* humans break rules for pro-social reasons. We hope that this informs the development of future AI agents, by modelling their decision-making accordingly. To do this, we introduced a new ethical dilemma called the vaccination strategy dilemma. Using this dilemma, we explored the relationship between external stakeholder utilities and PSRB. We found that the stakeholder utilities have a significant effect on human PSRB behaviour at least in the vaccine strategy dilemma. Also, the experiment proved that PSRB behaviour could not be categorised as a complete utilitarian behaviour. However, more research is needed to understand the relationship of PSRB with other external variables and internal variables mentioned in this paper.

# 9. Ethics Statement

The presented study has been granted exemption from requiring ethics approval by the Ethics Committee of the authors' university, University College Dublin, Ireland, under the protocol number UCD HREC-LS, Ref.-No.: LS-E-21-54-Ramanayake-Nallur. The study has been granted exemption as it included an anonymous survey that did not involve identifiable data, or any vulnerable groups. All participants voluntarily participated in the study, thus agreeing with the terms and conditions of the platform Amazon Mechanical Turk. All procedures performed were per the ethical standards of the institutional and national research committee (UCD HREC-LS, Ref.-No.: LS-E-21-54-Ramanayake-Nallur) and with the 1964 Helsinki declaration and its later amendments or comparable ethical standards.